\documentclass[useAMS,usenatbib]{mn2e}
\usepackage{times}
\usepackage{graphicx}

\usepackage{amssymb}

\usepackage[T1]{fontenc}
\usepackage{aecompl}
\usepackage{epstopdf}

\renewcommand{\vec}{\mathbf}

\begin{document}

\title[He\,{\sc II} intensity mapping]{Looking for Population III stars with He\,{\sc II} line intensity mapping}

\author[E. Visbal et al.]{Eli Visbal$^1$\thanks{visbal@astro.columbia.edu} \thanks{Columbia Prize Postdoctoral Fellow in the Natural Sciences}, Zolt\'{a}n Haiman$^1$, Greg L. Bryan$^1$ \\ $^1$Department of Astronomy, Columbia University, 550 West 120th Street, New York, NY, 10027, U.S.A. }

\maketitle

\begin{abstract}
Constraining the properties of Population III (Pop III) stars will be very challenging because they reside in small galaxies at high redshift which will be difficult to directly detect. In this paper, we suggest that intensity mapping may be a promising method to study Pop III stars. Intensity mapping is a technique proposed to measure large-scale fluctuations of galaxy line emission in three dimensions without resolving individual sources. This technique is well suited for observing many faint galaxies because it can measure their cumulative emission even if they cannot be directly detected.
We focus on intensity mapping of He\,{\sc II} recombination lines. These lines are much stronger in Pop III stars than Pop II stars because the harder spectra of Pop III stars are expected to produce many He\,{\sc II} ionizing photons. 
Measuring the He\,{\sc II} 1640~\AA{} intensity mapping signal, along with the signals from other lines such as Ly$\alpha$, H$\alpha$, and metal lines, could give constraints on the initial mass function (IMF) and star formation rate density (SFRD) of Pop III stars as a function of redshift.
To demonstrate the feasibility of these observations, we estimate the strength of the Pop III He\,{\sc II} 1640 \AA{} intensity mapping signal from $z=10-20$. We show that at $z\approx10$, the signal could be measured accurately by two different hypothetical future instruments, one which cross-correlates
 He\,{\sc II} 1640 \AA{} with CO(1-0) line emission from galaxies and the other with 21 cm emission from the intergalactic medium (IGM).
\end{abstract}


\section{Introduction}
There has been much theoretical work examining Pop III stars assembled from primordial gas in the early Universe. The first Pop III stars are expected to form in $(10^{5}-10^6) M_\odot$ minihalos that undergo molecular hydrogen cooling \citep{1996ApJ...464..523H,1997ApJ...474....1T,2002Sci...295...93A}. As the Universe evolves, stars produce a cosmological background of Lyman-Werner (LW) radiation which dissociates molecular hydrogen, eventually preventing additional star formation in low--mass minihalos \citep{1997ApJ...476..458H,2001ApJ...548..509M,2007ApJ...671.1559W,2008ApJ...673...14O,2011MNRAS.418..838W,2014MNRAS.445..107V}. Star formation is then restricted to larger halos, with masses near that of atomic cooling halos (i.e. halos with virial temperature above $T_{\rm vir} \approx 10^4 ~\rm{K}$). Three-dimensional hydrodynamical simulations have been utilized to understand the formation of Pop III stars both in minihalos and atomic cooling halos \citep[e.g.][]{2010MNRAS.403...45S,2011ApJ...727..110C,2012ApJ...760L..37H,2012MNRAS.424..399G,2012ApJ...745...50W,2014ApJ...781...60H}. Although these simulations cannot make predictions for the properties of Pop III stars, they have provided estimates for the sizes of the gas clumps forming as a result of fragmentation. The results suggest that the IMF of Pop III stars may have been top--heavy, but the IMF remains highly uncertain.

Unfortunately, it will be extremely challenging to test predictions for Pop III stars by observing them directly. This is because they are predicted to form in very small halos at great distances. In fact, even the James Webb Space Telescope (\emph{JWST}) \citep{2006SSRv..123..485G} may not be able to detect Pop III stars in atomic cooling halos at the epoch of reionziation ($z\sim10$) \citep{2011ApJ...740...13Z}. Pop III stars in minihalos at earlier times will be even more difficult to observe.  While it will be very interesting to study larger systems at high redshift, these galaxies are likely to have been enriched with metals from previous generations of star formation and therefore may not host significant Pop III star formation.

In this paper, we propose a new technique to constrain the properties of Pop III stars, intensity mapping of helium recombination lines. Intensity mapping is a technique to measure the large-scale three-dimensional clustering of galaxies by observing emission line fluctuations as a function of wavelength and position on the sky (\citealt{2010JCAP...11..016V,2011JCAP...08..010V}, see also \citealt{2008A&A...489..489R}), 
and has been proposed for a number of different lines \citep{2011ApJ...728L..46G,2011ApJ...741...70L,2012ApJ...745...49G,2013ApJ...763..132S,2013ApJ...768..130G,2014ApJ...786..111P}. Intensity mapping does not require resolving individual sources, instead it measures the cumulative line emission from \emph{all} galaxies in coarse three-dimensional pixels. This makes intensity mapping a promising technique to study the small systems that host Pop III star formation. Even though individual Pop III galaxies may be undetectable, their large numbers will result in a cumulative signal that could be measured through intensity mapping.

We focus on He\,{\sc II} lines (and in particular He\,{\sc II} 1640 \AA) because, compared to metal-enriched (``Pop II'') stars, the hard spectra of Pop III stars are expected to produce many more He\,{\sc II} ionizing photons relative to H\,{\sc I} ionizing photons, leading to strong recombination lines \citep{2002A&A...382...28S,2003A&A...397..527S}. For this reason, Pop III stars are likely to dominate the He\,{\sc II} intensity mapping signal (see Section 5). Studying Pop III stars with He\,{\sc II} lines has been considered in the context of individual sources 
\citep{TS2000,2001ApJ...553...73O,2011ApJ...736L..28C,2014arXiv1412.3845C,2013A&A...556A..68C}, but as mentioned above, it will be very difficult to detect the smallest Pop III sources, even with future telescopes. He\,{\sc II} lines have also been used to search for Pop III stars at lower redshifts, down to $z\sim 4$ \citep[e.g.][]{Dawson+2004,2006Natur.440..501J,2012MNRAS.419..479E}, with current results suggesting that the contribution of Pop III stars to the total SFRD is at most at the percent level \citep{Zheng+2013}.
Measuring the He\,{\sc II} intensity mapping signal, along with the intensity mapping signal of other lines such as Ly$\alpha$, H$\alpha$, and metal lines, will constrain the amount of Pop III star formation as a function of redshift and the hardness of their composite spectra which depends on the IMF.

This paper is structured as follows. In Section 2, we give a brief review of the intensity mapping technique. We discuss how the intensity mapping signal can be used to constrain the properties of Pop III stars in Section 3.  In Sections 4, 5, and 6, we estimate the strength of the He\,{\sc II} 1640 \AA{} intensity mapping signal from Pop III stars, Pop II stars, and quasars, respectively. In Section 7, we describe two hypothetical future intensity mapping experiments, one which cross-correlates He\,{\sc II} 1640 \AA{} with CO(1-0) (2610 $\mu$m) emission and the other with 21 cm emission from the IGM. As we discuss below, cross-correlation is necessary to remove contamination from other emission lines originating from galaxies at different redshifts. Finally, we discuss our results and conclusions in Section 8. Throughout we assume a $\Lambda$CDM cosmology consistent with the latest constraints from \emph{Planck} \citep{2014A&A...571A..16P}: $\Omega_\Lambda=0.68$, $\Omega_{\rm m}=0.32$, $\Omega_{\rm b}=0.049$, $h=0.67$, $\sigma_8=0.83$, and $n_{\rm s} = 0.96$.

\section{Intensity mapping review}
Here we review technical details of the intensity mapping technique. For a more thorough discussion see \cite{2010JCAP...11..016V,2011JCAP...08..010V}.
Intensity mapping requires measuring fluctuations in line intensity as a function of position on the sky and frequency, corresponding to a three-dimensional position in redshift + sky position space. The total intensity mapping fluctuation signal for a particular line, after removing the spectrally smooth components including galaxy continua and foreground/background radiation, is given by
\begin{equation}
\Delta S_{\rm tot}(\vec{r}) = \Delta S(\vec{r}) + \sum_{i} \Delta S_{{\rm cont},i}(\vec{r}_i') + \Delta S_{\rm Noise},
\end{equation}
where $\Delta S$ is the signal from the target line at the target redshift + sky position pixel $\vec{r}$, the $\Delta S_{{\rm cont}, i}$'s are the contaminating emission lines appearing at the target frequency from unrelated galaxies at different redshifts, and $\Delta S_{\rm Noise}$ is the detector noise. Assuming galaxies are found in dark matter halos which trace the large-scale cosmological density field, spatial fluctuations in galaxy line intensity are given by 
\begin{equation}
\Delta S = \bar{S} \bar{b} \delta + \Delta S_{\rm Poiss}, 
\end{equation}
where $\bar{S}$ is the mean line signal (proportional to the total line emission per volume), $\bar{b}$ is the mean line luminosity-weighted galaxy bias, $\delta$ is the cosmological dark matter overdensity, and $\Delta S_{\rm Poiss}$ is the variation due to Poisson fluctuations in the number of galaxies at fixed $\delta$. Throughout we approximate the mean bias as 
\begin{equation}
\bar{b} = \frac{\int dM b(M) L(M) \frac{dn}{dM}}{\int dM L(M) \frac{dn}{dM}},
\end{equation}
where $L(M)$ is the line luminosity from halos of mass $M$, $b(M)$ is the Sheth-Tormen halo bias, and $\frac{dn}{dM}$ is the halo mass function \citep{1999MNRAS.308..119S}. 

As discussed in \cite{2010JCAP...11..016V}, it is possible to isolate emission from a particular target redshift by cross-correlating different lines from the same volume. The cross-correlation function is defined by
\begin{equation}
\xi_{1,2}(\vec{r}) = \left \langle \Delta S_1(\vec{x}) \Delta S_2(\vec{x}+\vec{r})  \right \rangle 
\end{equation}
where the subscripts denote different emission lines. The cross power spectrum is defined as the Fourier transform of the cross-correlation function 
\begin{equation}
P_{1,2}(\vec{k}) = \int d^3 \vec{r} \xi_{1,2}(\vec{r}) e^{i \vec{k} \cdot \vec{r}},
\end{equation}
and (ignoring redshift space distortions) is given by 
\begin{equation}
\label{cc_eqn}
P_{1,2}(\vec{k}) = \bar{S}_1 \bar{b}_1  \bar{S}_2 \bar{b}_2 P(\vec{k}) + P_{\rm Poiss},
\end{equation} 
where $P(\vec{k})$ is the matter power spectrum and $P_{\rm Poiss}$ is the (constant) contribution from the Poisson fluctuations. The error on a measurement of the power spectrum for one k-mode is given by
\begin{equation}
\delta P_{1,2}^2 = \frac{1}{2} \left (P_{1,2}^2 + P_{\rm tot, 1} P_{\rm tot, 2} \right ),
\end{equation}
where $P_{\rm tot, 1}$ and $P_{\rm tot, 2}$ are the total power spectra for each data set being cross-correlated (see Appendix A of \cite{2010JCAP...11..016V} for a derivation of this formula). This is given by
\begin{equation}
\label{total_power}
P_{\rm tot, 1} = P_1+ P_{\rm Noise} + \sum_{ i}{P_{{\rm cont},i}},
\end{equation}
where $P_1$ is the power spectrum of the target line, $P_{\rm Noise}$ is the contribution from detector noise, and the $P_{{\rm cont},i}$'s are the power spectra from contaminating lines at different redshifts. These lines come from galaxies whose positions are assumed to be uncorrelated with those of the target galaxies because they are at different redshifts separated by very large distances (see below).

\section{Interpreting the intensity mapping signal}
In order to extract the most information from intensity mapping experiments it will be useful to measure the signal from a number of different lines. By taking the ratios of different lines' clustering cross power spectra (i.e. the first term in Eqn. \ref{cc_eqn}), one can find $\bar{S}\bar{b}$ for each line as a function of redshift. Additionally, the mean signal and bias could potentially be separated by examining the angular dependence of the power spectrum due to redshift space distortions.

To study Pop III stars it will be useful to measure the intensity mapping signal of He\,{\sc II} 1640 \AA{}, Ly$\alpha$, H$\alpha$, and a number of different metal lines. As discussed below, two possible (but most likely sub-dominant) non-Pop III contributions to the He\,{\sc II} signal come from Wolf-Rayet (WR) stars and (mini--) quasars. By measuring the intensity mapping signal of metal lines expected from WR stars or quasars in addition to He\,{\sc II}, it will be possible to verify that Pop III stars dominate the He\,{\sc II} signal and potentially subtract other contributions. It may also be possible to use metal lines to estimate the total H$\alpha$ or Ly$\alpha$ emission from Pop II stars and subtract it away, isolating the Pop III signal. One could then infer the total He\,{\sc II} 1640 \AA{} and H$\alpha$ signal from Pop III stars as a function of redshift. The ratio of these lines is a measure of the hardness of Pop III spectra which depends on the IMF (see \citealt{2001ApJ...553...73O} and fig.~5 in \citealt{2003A&A...397..527S}).  The Ly$\alpha$ line could be combined in a similar way, but would yield more model dependent results due to 
radiative transfer effects in the IGM that strongly modify the
apparent Ly$\alpha$ lineshape (e.g. \cite{ZH2002,Santos2004,Dijkstra2014}). The He\,{\sc II} signal
as a function of redshift would give information about how the SFRD of
Pop III stars evolves over cosmic time.

\section{He\,{\sc II} 1640 \AA{} line signal from Pop III stars}
In this section, we estimate the strength of the signal from Pop III stars to demonstrate the feasibility of He\,{\sc II} 1640 \AA{} intensity mapping. 
We assume that the line luminosity is given by
\begin{equation}
\label{line_lum}
L_{\rm HeII} = R \times {\rm SFR} \times \left ( 1-f_{\rm esc} \right ),
\end{equation} 
where SFR is the star formation rate, $R$ is a constant that depends on the IMF (and the metallicity if the stars are not metal-free), and $f_{\rm esc}$ is the escape fraction of HeII--ionizing (>54.4eV) photons,
which we assume to be much less than unity. For constant star formation, zero metallicity, and a Salpeter IMF with a mass range of $50-500$, $1-500$, or $1-100~M_\odot$, $R$ has a value of $1.6\times10^8$, $2.5\times10^7$,
or  $4.6 \times10^6~ L_\odot /(M_\odot {\rm yr^{-1}})$, respectively \citep{2003A&A...397..527S}.

\begin{figure} 
\includegraphics[width=88mm]{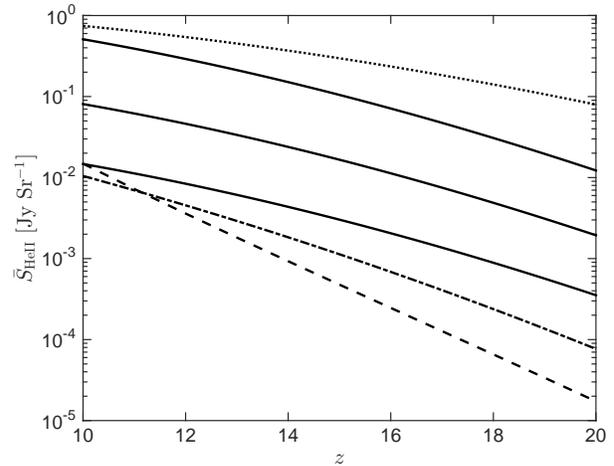}
\caption{Mean He\,{\sc II} 1640 \AA{} specific surface brightness versus redshift (corresponding to an observed wavelength of $1640$ \AA{} $\times (1+z)$) for different Pop III IMFs and $f_*=0.1$. The IMFs are Salpeter with mass ranges of $(1-100)M_\odot$ (bottom solid curve), $(1-500)M_\odot$ (middle solid curve), and $(50-500)M_\odot$ (top solid curve).  The dotted curve is the signal from minihalos assuming no LW background (assuming a Salpeter IMF with mass range $(50-500)M_\odot$). The dot-dashed curve is the upper limit from (Pop II) WR stars discussed in Section 5 and the dashed curve is the contribution from quasars estimated in Section 6. \label{signal_plot} }
\end{figure}

\begin{figure} 
\includegraphics[width=88mm]{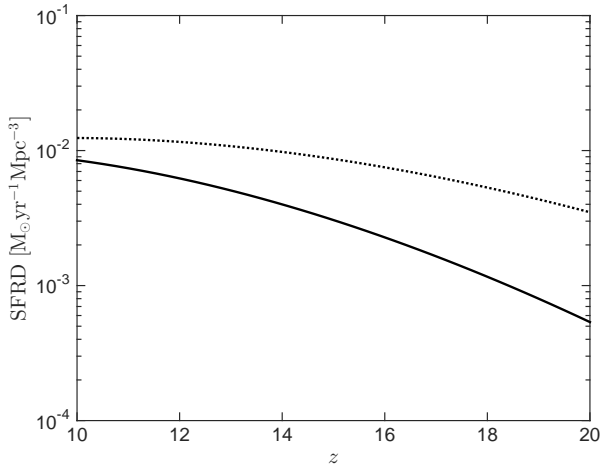}
\caption{ The SFRD corresponding to the He\,{\sc II} specific surface brightness plotted in Figure \ref{signal_plot} for atomic cooling halos (solid curve) and minihalos (dotted curve). Note that we have ignored the negative feedback on minihalos due to LW radiation.  \label{sfrd_plot} }
\end{figure}

In the examples discussed below, we consider He\,{\sc II} intensity mapping at redshifts $z > 10$. We note that this is a conservative choice, as Pop III stars could exist at redshifts as low as $z\sim 3-4$ \citep{2006Natur.440..501J,2007ApJ...659..890W,2011Sci...334.1245F} potentially producing a stronger signal, extending down to lower observed wavelengths.
We assume that the LW background is strong enough to suppress star formation in minihalos and focus on atomic cooling halos (but will address a possible extra contribution from minihalos below). For dark matter halos above the atomic cooling threshold we approximate the SFR as
\begin{equation}
\label{SFR_eqn}
{\rm SFR} = f_* \frac{\Omega_{\rm b}}{\Omega_{\rm m}} \frac{M} {\epsilon_{\rm duty} t_{\rm H}(z)},
\end{equation}
where  $M$ is halo mass, $f_*$ is the star formation efficiency, $\epsilon_{\rm duty}$ is the duty cycle, and $t_{\rm H}(z)$ is the Hubble time. To compute the total signal, we assume that Pop III stars are formed in halos with $M=(1-2) M_{\rm cool}$, where $M_{\rm cool}$ is the atomic cooling mass ($M_{\rm cool}=3\times 10^7 ((1+z)/11)^{1.5} M_\odot$ corresponding to $T_{\rm vir}=10^4$~K for neutral gas). Much larger halos are likely to have been enriched with metals from stars formed in their progenitor halos.  Given these assumptions, the mean Pop III He\,{\sc II} signal (i.e. the specific surface brightness, in units of $\rm{ergs~s^{-1}~cm^{-2}~Hz^{-1}~sr^{-1}}$ and proportional to line luminosity per volume of space) is 
\begin{equation}
\label{S_eqn}
\bar{S}_{\rm HeII} = \int_{M_{\rm cool}}^{2M_{\rm cool}} dM \frac{L_{\rm HeII}(M)}{4\pi D_L^2} \epsilon_{\rm duty} \frac{dn}{dM} \tilde{y} D_A^2,
\end{equation}
where  $D_L$ is the cosmological luminosity distance, $D_A$ is the angular diameter distance, and $\tilde{y}=\frac{d\chi}{d\nu}$ is the corresponding comoving distance ($\chi$) per change in observed frequency. Note that the mean signal does not depend on $\epsilon_{\rm duty}$. We plot the signal as a function of redshift for several different IMFs and $f_*=0.1$ in Figure \ref{signal_plot}.  For comparison, we also plot the signal for Pop III stars in minihalos assuming no LW background. We compute this using Eqn. \ref{S_eqn}, but replace the atomic cooling mass with the cooling mass for molecular cooling, $M_{\rm H_2} = 2.5 \times 10^5 \left (\frac{1+z}{26} \right )^{-3/2} M_\odot$ \citep{2001ApJ...548..509M,2007ApJ...671.1559W,2008ApJ...673...14O}.  Note that not including LW feedback results in a large overestimate below the redshift where this effect becomes important. In Figure \ref{sfrd_plot} we plot the SFRD of Pop III stars corresponding to the He\,{\sc II} specific surface brightness in Figure \ref{signal_plot}. We note that for Pop III stars in atomic cooling halos, the SFRD is within roughly an order of magnitude of that computed in \cite{2009ApJ...694..879T} when they assume multiple Pop III stars can form in each dark matter halo.

By using a constant duty cycle, combined with a snapshot of the halo mass function (rather than using the halo formation rate) Eqns. \ref{SFR_eqn} and \ref{S_eqn} give a rough approximation of $\bar{S}_{\rm HeII}$ and the corresponding Pop III SFRD.
To test this approximation, we estimate the SFRD (which is proportional to $\bar{S}_{\rm HeII}$) with a different approach and compare the results. Numerical simulations show that at $z\sim10$, the rate at which halos cross the atomic cooling threshold per volume is $\frac{dn_{\rm cool}}{dz}\sim 4  ~ \rm{Mpc}^{-3}$ \citep{2014MNRAS.445.1056V}. Assuming $f_*=0.1$ of the gas in these halos forms Pop III stars over a time period $t_{\rm SF}$ and that $\frac{dn_{\rm cool}}{dz}$ does not evolve rapidly over this time scale, the Pop III star formation rate density is given by ${\rm SFRD} = \frac{dn_{\rm cool}}{dz} \frac{dz}{dt}  \left ( f_* \frac{\Omega_{\rm b}}{\Omega_{\rm m}} M_{\rm cool} \right )$. The SFRD computed this way is $\sim 3.5$ times greater than that implied by Eqns. \ref{SFR_eqn} and \ref{S_eqn} (this increase is attributable to the rapid halo formation rate).
Thus, the actual fraction of gas that is required to form stars to reach a particular $\bar{S}_{\rm HeII}$ may be less than $f_*$ in Eqn. \ref{SFR_eqn}.

To check that our SFR prescription and fiducial choice of $f_*=0.1$ are reasonable, we compare the amount of ionizing photons produced at $z\sim10$ to that required to maintain reionization. The recombination rate per volume in a completely ionized region is
\begin{equation}
B = \alpha_{\rm B} n_{\rm IGM}^2 C,
\end{equation}
where $\alpha_{\rm B}=2.6\times 10^{-13} {\rm cm^3 s^{-1}}$ is the case B recombination coefficient, $n_{\rm IGM}$ is the mean hydrogen number density of the intergalactic medium (IGM), and $C=\langle n_{\rm IGM,HII}^2 \rangle/\langle n_{\rm IGM,HII} \rangle^2$ is the clumping factor of the ionized gas in the IGM. 
The rate of ionizations per volume from star formation is given by ${\rm SFRD} \times Q(H) \times f_{\rm esc, HI}$, where $Q(H)$ is the number of ionizing photons per SFR and $f_{\rm esc, HI}$ is the escape fraction of H\,{\sc I} ionizing photons. 
We compute the SFRD using Eqn. \ref{SFR_eqn} and integrating the halo mass function. We assume that Pop III stars with a Salpeter IMF and mass range of (50-500)$M_\odot$ form in halos with $M=(1-2)M_{\rm cool}$ and that Pop II stars with a metallicity of $Z=0.0004$ and a Salpeter IMF with mass range (1-100)$M_\odot$ form in larger halos. Using the values of $Q(H)$ from \cite{2003A&A...397..527S}, $C=3$ \citep{2012MNRAS.427.2464F}, $f_*=0.1$, and $f_{\rm esc, HI} =0.03$, we find that the rate of recombination equals the rate of ionization if the IGM were ionized. Thus, our fiducial choice of $f_*=0.1$ appears reasonable. 

\section{Pop II Wolf-Rayet signal}
The most significant source of He\,{\sc II} recombination lines from Pop II stars is likely to come from WR stars. Here we show that this component is unlikely to dominate the signal at high-redshift. It is difficult to estimate the WR signal from population synthesis models, so instead we utilize observational results.  \cite{2013A&A...556A..68C} compiled a sample of high-redshift galaxies ($z = 2-4.6$) with broad He\,{\sc II} lines consistent with those expected from WR stars. For each galaxy, they compare the SFR estimated from the broad-band SED to the He\,{\sc II} 1640 \AA{} luminosity and find the best fit linear relation
\begin{equation}
\label{wr_ul}
L_{\rm HeII, WR} = 3.5 \times 10^{39}\left ( \frac{\rm SFR}{\rm M_\odot ~yr^{-1}} \right ) {\rm ergs~s^{-1}}.
\end{equation}
This can be viewed as an upper limit on the WR contribution since most galaxies' spectra do not show strong He\,{\sc II} spectral features. 

This relation is consistent with the empirical estimates of $Q(\rm{HeII})/Q(\rm{H})$ (where $Q$s are ionization rates) inferred from the He\,{\sc II} 4686\AA{} to H $\beta$ line ratios in extragalactic HII regions discussed in \cite{2003A&A...397..527S} (see their section 6.3, see also \citealt{2000ApJ...531..776G}). We use the stellar models of \cite{2003A&A...397..527S} to translate these estimates into He\,{\sc II} 1640 \AA{} luminosity per SFR. Assuming a Salpeter IMF with a mass range of $M=(1-100)~M_\odot$ and a constant SFR, we find that for metallicity $Z= 10^{-3}$,
\begin{equation}
L_{\rm HeII, WR} \sim (1.4-5.5) \times 10^{39} \left ( \frac{\rm SFR}{\rm M_\odot ~yr^{-1}} \right ) {\rm ergs~s^{-1}}.
\end{equation}
The observations indicate that at higher metallicity the WR He\,{\sc II} signal is much lower.

Assuming that Pop II stars are found in dark matter halos above 2$M_{\rm cool}$  and using Eqn. \ref{wr_ul} and Eqn. \ref{SFR_eqn}, we plot the upper limit of the WR contribution to the He\,{\sc II} 1640 \AA{} intensity mapping signal in Figure \ref{signal_plot}. The Pop III component dominates at all redshifts for top heavy IMFs (i.e. Saltpeter with mass limits of $(50-500)M_\odot$ or $(1-500)M_\odot$); and it remains dominant at redshifts well above $z>10$ for a usual Salpeter IMF (with $(1-100)M_\odot$).

\section{Quasar signal}
The spectra of quasars are harder than stellar sources and thus can also efficiently ionize He\,{\sc II}. Here we estimate their contribution to the He\,{\sc II} 1640 \AA{} intensity mapping signal. To calculate the line strength, we use eqn. 2 from \cite{2001ApJ...553...73O}, which assumes a spectral template from \cite{1994ApJS...95....1E} normalized such that the total luminosity corresponds to the Eddington limit. This gives an He\,{\sc II} 1640 \AA{} luminosity of
\begin{equation}
L_{\rm HeII, quasar} = 10^{41} \left( \frac{M_{\rm BH}}{10^5~M_\odot} \right ) \rm{ergs~s^{-1}},
\end{equation}
where $M_{\rm BH}$ is the mass of the black hole powering the quasar (or mini-quasar). We assume that black hole mass relates to halo mass as
\begin{equation}
\label{bh_eqn}
M_{\rm BH} = 3 \times 10^9 \left (\frac{M}{5 \times 10^{12} ~M_\odot}  \right )^{5/3} \left ( \frac{1+z}{7} \right ) M_\odot,
\end{equation}
and compute the mean intensity mapping signal from the halo mass function. We plot the average signal versus redshift in Figure \ref{signal_plot} assuming that the atomic cooling mass is the minimum halo mass hosting quasars and that they have a duty cycle of 0.05 (a high duty cycle compared to low-redshift quasars is possible due to the increased frequency of major mergers at high redshift, \citealt{2014CQGra..31x4005T}). For this choice of duty cycle, Eqn. \ref{bh_eqn} implies that the density of active quasars hosting $\sim 3 \times 10^9 M_\odot$ or larger black holes is $\sim 10^{-9} {\rm Mpc^{-3}}$, which is consistent with observations \citep{2006NewAR..50..665F}. The mass and redshift scaling are expected for feedback-limited accretion \citep{2013fgu..book.....L}.
We find that the Pop III signal dominates for Saltpeter IMFs with mass limits of $(50-500)M_\odot$ or $(1-500)M_\odot$ at $z>10$. 

\section{Future Observations}
In this section, we discuss the capability of a hypothetical futuristic instrument to measure the He\,{\sc II} 1640 \AA{} intensity mapping signal. We consider a space-based instrument with a 2 m dish and background-limited sensitivity. The background radiation at the relevant wavelengths is dominated by Zodiacal light and faint stars. We use fig.~1 of \cite{1998A&AS..127....1L} to determine its strength. We assume that the instrument can simultaneously measure spectra of 100 separate 4 arcmin$^2$ pixels with spectral resolution $\nu/\Delta \nu = 1000$. This design represents an instrument which is not feasible with current technology, but could potentially be built in the near future.

We consider two separate examples, one cross-correlating He\,{\sc II} 1640 \AA{} with CO(1-0) emission from galaxies and the other cross-correlating with 21 cm emission from the IGM. For both examples, we compute the total signal-to-noise of the cross power spectrum defined as
\begin{equation}
\label{sn_eqn}
\frac{S}{N} = \sqrt{\sum_{ \vec{k}-{\rm modes}} \frac{P_{1,2}^2}{\delta P_{1,2}^2}},
\end{equation}
where the sum is over the available k-modes, determined by the dimensions and spatial resolution of the survey. We note that modes at the redshifts and scales we consider are in the linear regime for the target galaxies and thus are statistically independent. 
Ignoring redshift-space distortions (described below) and assuming the value of the matter power spectrum can be estimated accurately from constraints on cosmological parameters, $\frac{S}{N}$ corresponds to the uncertainty on $\bar{S}_1 \bar{b}_1 \bar{S}_2 \bar{b}_2$ measured with an inverse variance--weighted average.  We note that for the observations described below, the Poisson component of the signal is negligible compared to the clustering component for the scales we consider.
 
A large source of noise in the He\,{\sc II} cross power spectrum comes from emission in numerous contaminating lines (e.g. H$\alpha$) from foreground and background galaxies (the third term on the right hand side of Eqn. \ref{total_power}). To estimate this noise, we assume that galaxy line luminosity is proportional to SFR and calibrate the strength of the emission based on the typical starburst galaxy NGC 7714. We use Table~3A in \cite{1995ApJ...439..604G} and Table~5 in \cite{2006A&A...457...61R} to determine the relative strength of $\sim 50$ lines with rest frame wavelengths between 3727\AA{} and 2.2$\mu$m.  Contaminating galaxies are assumed to be hosted by dark matter halos with mass greater than $2M_{\rm cool}$ at $z>6$ and Eqn. \ref{SFR_eqn} is used to determine their SFR. For lower redshift, we assume that radiative feedback on the IGM from reionization increases the minimum virial temperature of halos hosting galaxies to $T_{\rm vir}= 10^5 $~K. We assign a SFR to these galaxies that is  proportional to halo mass, with a normalization that sets the total SFRD (determined by integrating the halo mass function) to
\begin{equation}
{\rm SFRD}(z) = \frac{a+bz}{1+(z/c)^d} \left ( h M_\odot {\rm yr^{-1} Mpc^{-3}} \right ),
\end{equation}
where $a=0.0118$, $b=0.08$, $c=3.3$, and $d=5.2$, which agrees with observations \citep{2006ApJ...651..142H}.

In order to increase the signal-to-noise of the cross power spectrum measurement, it will be useful to either mask or subtract away line emission from bright foreground galaxies in the He\,{\sc II} intensity map. We consider two different conditions for removing these contaminating lines. In the more conservative case, we assume that lines are removed if their strengths (calibrated with NGC 7714 as described above) are more than 5 times the detector noise. This effectively sets a maximum mass of contaminating foreground halos
(i.e. we impose an upper limit on the integral over halo mass for the contamination signal, analogous to Eqn.~\ref{S_eqn}).
In practice, it would be necessary to identify a few lines at a few different wavelengths from the same contaminating galaxy. We do not consider this procedure in detail, instead we assume galaxies with high signal-to-noise lines could be identified and removed. In the second, optimistic case, we assume that data from other telescopes could be used to mask or subtract line emission from bright foreground galaxies. In particular, we remove any galaxies which could be observed with the Large Synoptic Survey Telescope\footnote{http://www.lsst.org/} (\emph{LSST}) ``deep drilling'' fields. Galaxies with UV magnitude less than $m_{\rm AB}=28$ are removed. The UV luminosity is computed with the following relation \citep{1998ARA&A..36..189K}, 
\begin{equation}
\frac{\rm SFR}{[M_\odot~{\rm yr}^{-1}]}= 1.4\times10^{-28} \frac{L_{\nu}}{\rm ergs~s^{-1}~Hz^{-1}}.
\end{equation}

In our signal-to-noise calculations, we also include the linear change to the power spectrum from 
redshift-space distortions  \citep[see the Appendix of][]{2014ApJ...785...72G}, which effectively changes the bias from $\bar{b} \rightarrow \bar{b} + \cos^2{\theta}\frac{d\ln{D}}{d\ln{a}}$, where $D$ is the growth function, $a$ is the scale factor, and $\theta$ is the angle between $\vec{k}$ and the line of sight. This is modified slightly for background and foreground galaxies due to the varying relationship between distance and corresponding angle/observed frequency. We also note that the matter power spectrum can enter the non-linear regime for low-redshift foreground galaxies. We use the fit from \cite{2012ApJ...761..152T} to compute the increase in power due to non-linearity.  

\subsection{CO(1-0)-He\,{\sc II} cross-correlation}
For our first example, we consider a measurement of the CO(1-0) (2610 $\mu$m)-He\,{\sc II} 1640 \AA{} cross power spectrum. To estimate the strength of the CO signal, we assume that CO line emission comes from halos above 2$M_{\rm cool}$ (while halos below this mass host Pop III star formation) and the line luminosity is proportional to halo mass. We use the normalization from \cite{2011ApJ...741...70L},
\begin{equation}
L_{\rm CO(1-0)} = 2.8 \times 10^3 \frac{M}{10^8 M_\odot} L_\odot,
\end{equation}
and assume that CO emission comes from a fraction of halos corresponding to a duty cycle $\epsilon_{\rm duty} =0.1$. These assumptions give a mean signal of $\bar{S}_{\rm CO(1-0)}=4.5~ \rm{\mu K}$ ($15~{\rm Jy ~sr^{-1}}$) at $z=10$. This is similar to the results of other studies, but corresponds to optimistic estimates for a very uncertain signal. The true signal could be lower if, for example, low-mass galaxies near $2M_{\rm cool}$ do not have sufficient metals to produce a strong CO(1-0) line.

We examine the potential of the He\,{\sc II} 1640 \AA{} intensity mapping instrument described above and a separate CO experiment. We assume the CO experiment matches the 2 arcmin angular resolution of the He\,{\sc II} instrument and that both intensity maps have a total sky coverage of 400 deg$^2$ over the entire course of the observation (combined as a mosaic from many pointings on the sky). We also assume matching spectral resolution of $\nu/\Delta \nu=1000$ and that each observation spans a total bandwidth of $B=0.1\nu$. 

For the CO(1-0) survey, we consider an interferometer with detector noise
\begin{equation}
\sigma_{\rm N}  \approx \frac{T_{\rm sys}}{\sqrt{\Delta \nu t_{\rm field}}} \frac{1}{f_{\rm cover}},
\end{equation}
where the system temperature is taken to be $T_{\rm sys}=30~{\rm K}$ and $t_{\rm field}$ is the integration time per field of view (assuming the observation is a mosaic of many such fields). The covering factor $f_{\rm cover}= N_{\rm a} D^2/D_{\rm max}^2$, where $N_{\rm a}$ is the number of antennae, $D$ is the size of each antenna, and $D_{\rm max}$ is the maximum base line, which sets the angular resolution. Assuming $D=0.4 ~{\rm m}$ (corresponding to a 4$^\circ$ field of view at $z=10$), $D_{\rm max}=50~{\rm m}$ (corresponding to 2 arcmin resolution at $z=10$), $N_{\rm a}=10,000$, and a total integration time of 1 yr, we find $\sigma_{\rm N}  \approx 10~ \rm{\mu K}$ ($33~{\rm Jy ~sr^{-1}}$). We use this value of $\sigma_N$ for all of our calculations.  This represents a futuristic, second generation intensity mapping experiment. Note that this formula assumes uniform coverage in Fourier space and is a conservative estimate, somewhat fewer antennae may be needed for this sensitivity \citep{2011ApJ...741...70L}. The noise assumed here is intended as a rough estimate, future work will be required to optimize the design of CO intensity mapping instruments. For the He\,{\sc II} survey, we determine the noise by assuming that the entire observation consists of one year of integration time. 

\begin{figure} 
\includegraphics[width=88mm]{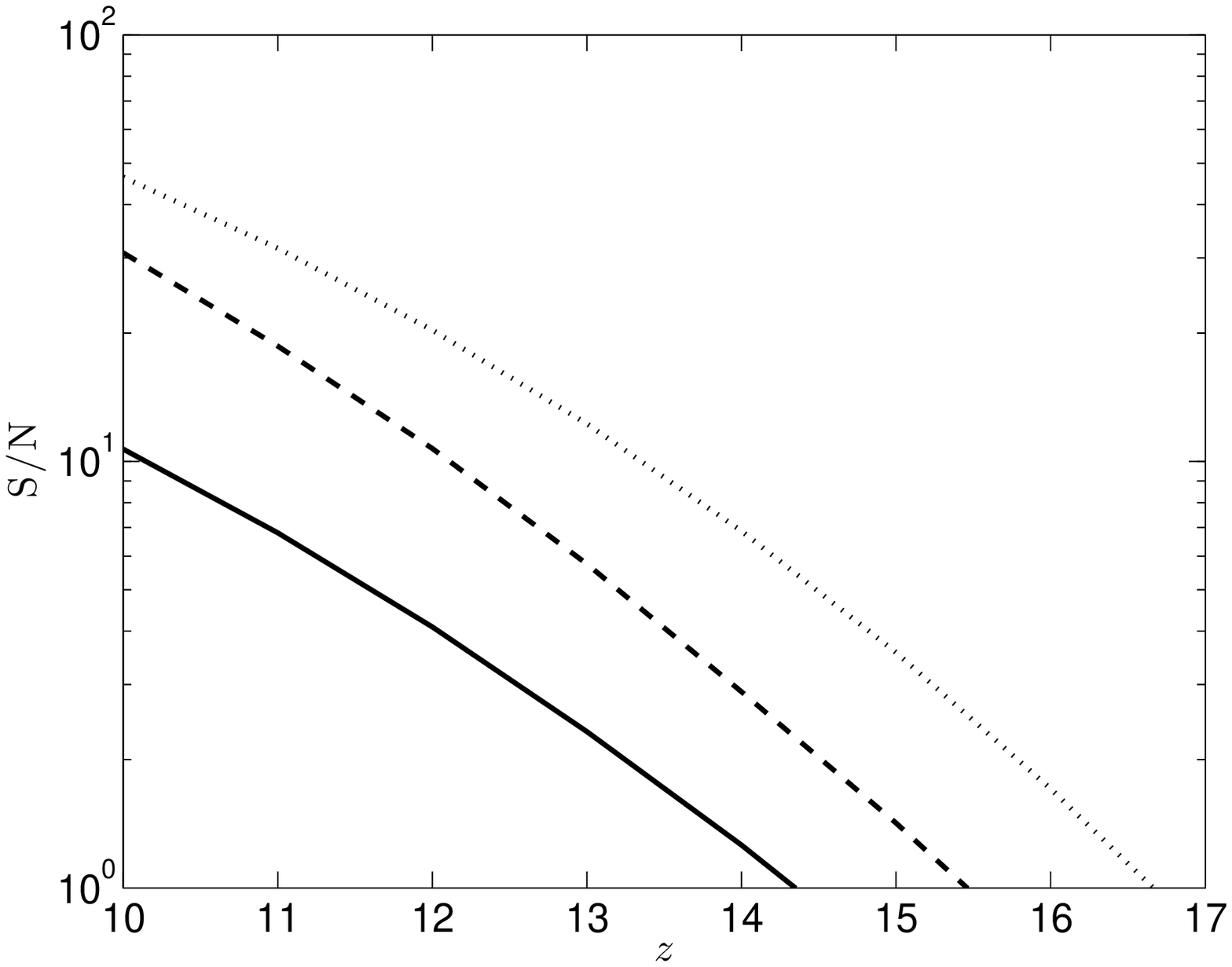}
\includegraphics[width=88mm]{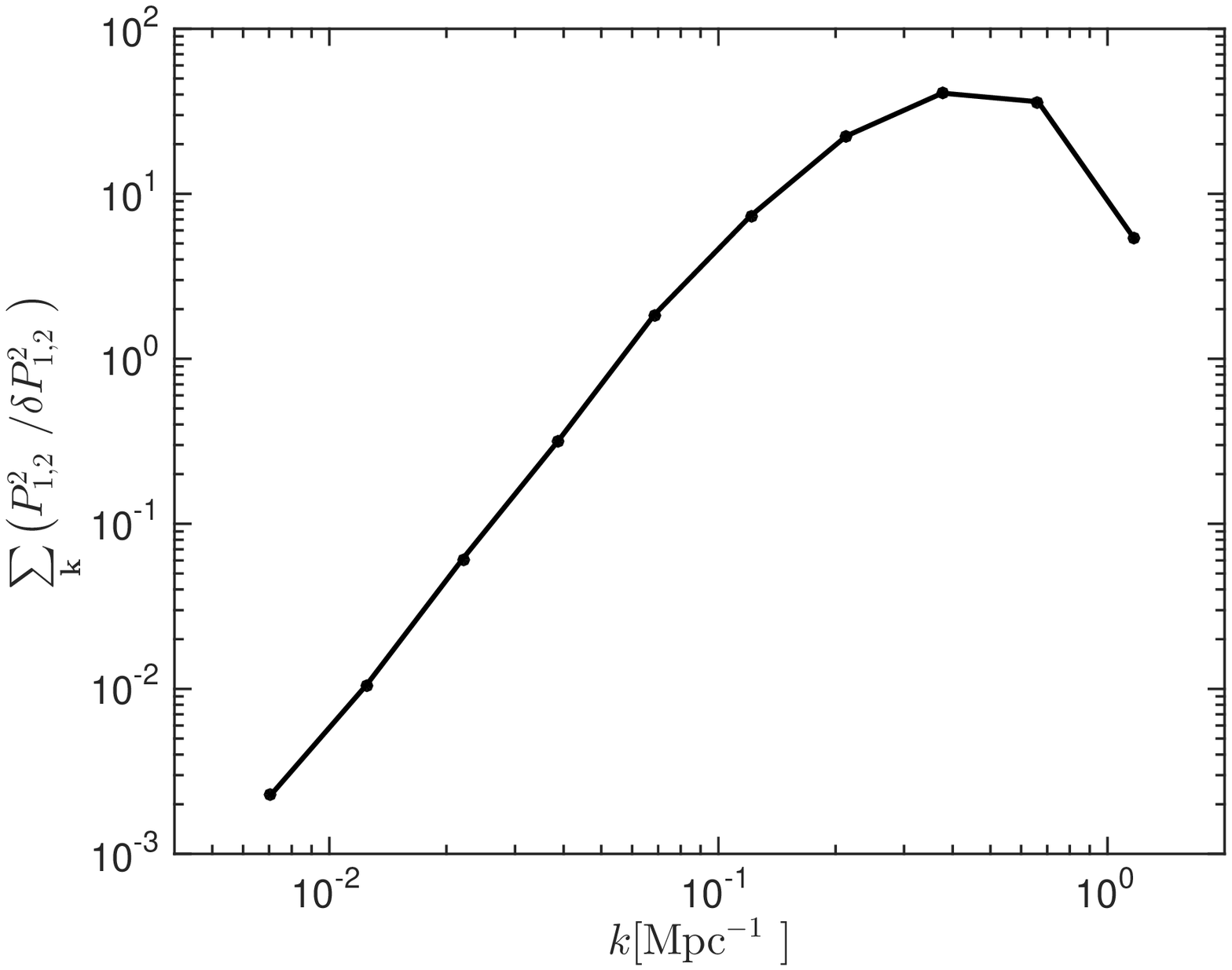}
\caption{Top panel: signal-to-noise for detecting the He\,{\sc II} 1640 \AA{}-CO(1-0) cross power spectrum for the hypothetical surveys described in Section 7.1  (see Eqn. \ref{sn_eqn}). We assume a Salpeter IMF with mass range of (50-500)$M_\odot$ (a mass range of $(1-500)M_{\odot}$ would result in a signal-to-noise $\sim6$ times lower). The solid and dashed lines are for our conservative and optimistic prescriptions for removing foreground galaxies. For reference, the dotted line is the signal-to-noise with no contaminating lines from different redshifts. Bottom Panel: The contribution to the signal-to-noise squared as a function of scale for the conservative contamination case at $z=10$. The contribution is the total from all modes in $|\vec{k}|$-bins. Most of the information comes from scales corresponding to $k\sim (0.2-0.7) ~ {\rm Mpc^{-1}}$.
 \label{SN_He_CO} }
\end{figure}

In Figure \ref{SN_He_CO}, we plot the signal-to-noise of the cross power spectrum for this measurement as a function of redshift assuming a Salpeter IMF with a mass range of (50-500)$M_\odot$. The signal-to-noise would be roughly $6$ times lower for a mass range of $(1-500)M_\odot$. The Pop III He\,{\sc II} 1640 \AA{} signal could potentially be accurately measured at $z=10$ and beyond.

\subsection{21 cm-He\,{\sc II} cross-correlation}
\begin{figure} 
\includegraphics[width=88mm]{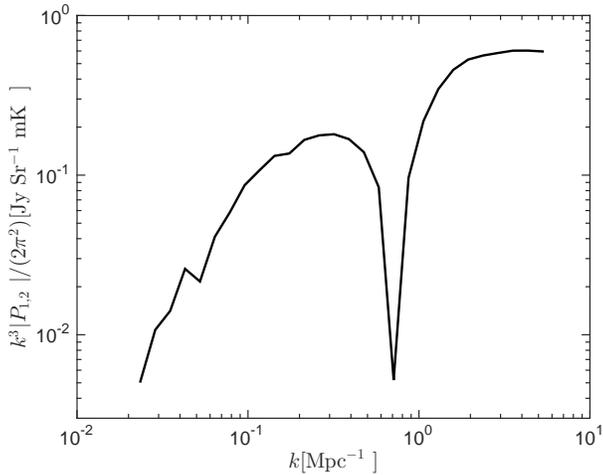}
\caption{\label{21 cm_power}  The absolute value of the He\,{\sc II} 1640 \AA{}-21 cm cross power spectrum at $z=10$. The sign of the cross power spectrum is negative for $k < 0.7 ~{\rm Mpc^{-3}}$, where the dip occurs (which is related to the typical size of ionized bubbles in the IGM). For Pop III stars, we have assumed a Salpeter IMF with a mass range of $(50-500)M_{\odot}$.  }
\end{figure}

Next we consider a future measurement of the He\,{\sc II} 1640 \AA{}-21 cm cross power spectrum. To estimate the value of the power spectrum we use the publicly available code 21cmFAST \citep{2011MNRAS.411..955M}. We compute the ionization field at $z=10$ assuming a minimum halo mass of ionizing sources equal to $M_{\rm cool}$ and an ionizing efficiency that sets a region as ionized if its collapsed fraction is greater than $1/12$.
This leads to a neutral fraction of $\bar{x}_{\rm H}=0.68$ at $z=10$.  Assuming the IGM temperature is much greater than CMB temperature and ignoring peculiar velocities, the 21 cm fluctuation signal is given by
 \begin{equation}
 \Delta T_{21}  = 30 x_{\rm H} \left (1 + \delta_{\rm b} \right) \sqrt{\frac{1+z}{10}} ~ {\rm mK},
 \end{equation}
where $x_{\rm H}$ is the neutral fraction as a function of position and the baryon overdensity, $\delta_{\rm b}$, is assumed to equal the total matter overdensity, $\delta$. We then compute the cross power spectrum by taking the discrete Fourier transform of $\Delta T_{21}$ and the He\,{\sc II} 1640 \AA{} intensity mapping signal, $\Delta S_{\rm HeII} = \bar{S}_{\rm HeII}\bar{b}\delta$. The $z=10$ power spectrum is shown in Figure \ref{21 cm_power}. Note that the HeII-21 cm cross power spectrum is not of the form of Eqn. \ref{cc_eqn}. However, the ratio of the cross power spectra can still be taken for different galaxy lines to give $\bar{S}\bar{b}$ as a function of redshift. For simplicity we do not consider redshift space distortions for the He\,{\sc II}-21 cm power spectrum.

\begin{figure} 
\includegraphics[width=88mm]{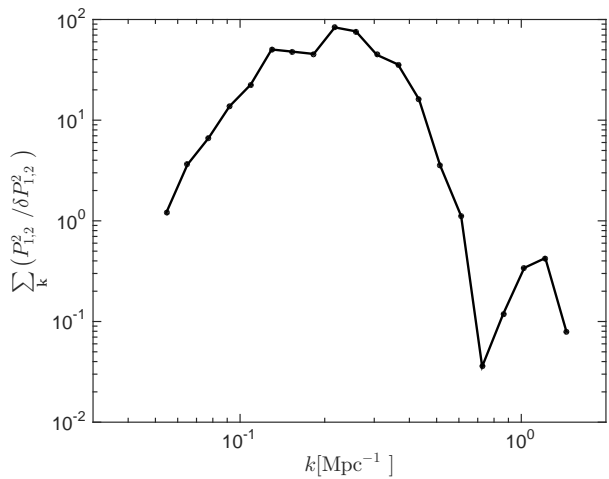}
\caption{ \label{21 cm_SN} The contribution to the signal-to-noise squared as a function of scale for the HeII-21 cm cross power spectrum at $z=10$. The total signal-to-noise is $\frac{S}{N}=21$. For Pop III stars we have assumed a Salpeter IMF with a mass range of $(50-500)M_{\odot}$. We use the optimistic consideration described in the text for contaminating line removal in the He\,{\sc II} intensity map.}
\end{figure}

We calculate the signal-to-noise of the cross power spectrum for the hypothetical He\,{\sc II} instrument described above and a next generation 21 cm experiment. For the 21 cm experiment, we consider a scaled up version of the Murchison Widefield Array\footnote{http://www.mwatelescope.org/} (\emph{MWA}). We assume an interferometer with 5000 tiles each with effective area $A_{\rm e} = 16~\rm{m}^2$. The tiles are distributed in a constant core up to a radius of 60 m and fall off at larger radii as $n_{\rm tile} \propto r^{-2}$. The layout determines the baseline distribution, which controls k-mode coverage in the direction perpendicular to the line of sight. The total 21 cm power spectrum (including detector noise) is given by  \citep{2006ApJ...653..815M} 
\begin{equation}
P_{\rm tot, 21}(\vec{k}) = P_{21}(\vec{k}) +  D_{\rm A}^2\tilde{y} \left ( \frac{\lambda^2 T_{\rm sys}}{A_{\rm e}} \right )^2 \frac{1}{tn(k_{\perp})},
\end{equation}
where $P_{21}$ is the cosmological 21 cm signal, $\lambda=(1+z) \times 21~{\rm cm}$, $t=3 ~{\rm yr}$ is the total observation time (1 yr total integration time is assumed for the He\,{\sc II} observation), and $n(k_{\perp})$ is the baseline density. The system temperature is dominated by the sky temperature (mostly from Galactic synchrotron emission) and given by $T_{\rm sys}=240 \left ( \frac{1+z}{9.5} \right )^{2.5} ~\rm{K}$ \citep{2008AJ....136..641R}. 
The total field of view of the observation is taken to be 1000 deg$^2$ and we assume that the angular resolution of the He\,{\sc II} instrument is 3 arcmin in this case to better match the 21 cm experiment. 
The total signal-to-noise at $z=10$, given by equation \ref{sn_eqn} is $S/N=21$, assuming the optimistic condition for removing foreground galaxies in the He\,{\sc II} map. We plot the contribution from different scales in Figure \ref{21 cm_SN}. 

\section{Conclusions}
In this paper, we argue that intensity mapping is a promising technique to constrain the properties of Pop III stars because it can measure the cumulative line emission from many sources that are too faint to be resolved individually. This is particularly useful for studying Pop III stars because they are expected to be found in very faint galaxies at high redshift which will be difficult to observe individually even with future telescopes such as \emph{JWST}. 

We focus on intensity mapping of He\,{\sc II} recombination lines (the 1640 \AA{} line in particular) because the hard spectrum predicted for Pop III stars makes it likely that their contribution will be dominant compared to Pop II stars or quasars (see Sections 5 and 6). Measuring the He\,{\sc II} 1640 \AA{}  intensity mapping signal as well the signal from other lines such as H$\alpha$, Ly$\alpha$ \citep{2014ApJ...786..111P,2013ApJ...763..132S}, and metal lines, will enable one to quantify the total Pop III He\,{\sc II} emission and the Pop III spectral hardness (i.e. the ratio of He\,{\sc II} to H ionizations) as a function of redshift. This will constrain the Pop III IMF and the evolution of the Pop III SFRD. 
Constraints on Pop III stars may also be improved with intensity mapping of H$_2$ cooling lines which are produced during star formation \citep{2013ApJ...768..130G}.
Future work is required to determine how to best combine intensity mapping of different emission lines to put the tightest possible constraints on Pop III star formation.

We estimated the Pop III He\,{\sc II} 1640 \AA{} intensity mapping signal from atomic cooling halos at $z=10-20$ because atomic cooling halos are likely to be brighter and at lower redshift than minihalos. However, it may be possible to detect the signal from minihalos as well. Their signal at $z\sim20$ could be nearly as high as the atomic cooling halo signal at $z\sim10$ if the LW background were very low and/or ineffective in sterilizing minihalos (see Figure \ref{signal_plot}).  We consider two examples of future intensity mapping observations. In the first, He\,{\sc II} 1640 \AA{} is cross-correlated with CO(1-0) emission from galaxies and in the second He\,{\sc II} is cross-correlated with 21 cm radiation from the IGM. Cross-correlation is necessary to remove contamination due to line emission from foreground and background galaxies. We find that in both examples the cross power spectrum can be measured with high signal-to-noise for Salpeter IMFs with mass ranges of $(50-500)M_\odot$ and $(1-500)M_\odot$ at $z\sim10$.

There are several caveats that could impact the effectiveness of this technique. First, there may be some degeneracy in the total He\,{\sc II} signal and spectral hardness for Pop III metal-free stars with a particular IMF and stars which have a very low metallicity and a slightly different IMF (see fig. 5 in \citealt{2003A&A...397..527S}). There may also be degeneracy for different functional forms of the Pop III IMF. However, in any case, a very hard spectrum is likely to be strong evidence of a top-heavy Pop III IMF. Throughout we have assumed that the escape fraction of ionizing photons is much less than unity. We regard this as a reasonable assumption.  If HeII-ionizing photons were to leak out of galaxies, they would still be absorbed in the local IGM and produce only a small reduction in our signal (note that the HeII recombination time in the IGM at $z>10$ is shorter than the Hubble time).  Additionally, we stress that our estimates for the contribution from quasars are very uncertain. If the quasar signal is much higher than we estimate in section 6, He\,{\sc II} intensity mapping may be a better probe of high-redshift quasars than Pop III stars. This would still be very valuable; the inferred presence of a population of faint undetected miniquasars could be studied by correlations with the X-ray observations.
Our estimate of the He\,{\sc II} signal is also only intended to be a rough estimate. If the true signal is much lower than the values we compute, it may be necessary to design significantly more ambitious instruments than those we describe for a high signal-to-noise detection. Despite all of the complications described above, because Pop III galaxies are so difficult to study individually, aside from observations of Supernovae \citep[e.g.][]{2005ApJ...629..615W,2006ApJ...637...80M}, He\,{\sc II} intensity mapping may be the most direct method of constraining the Pop III IMF and the SFRD evolution at $z\gtrsim 10$ in the foreseeable future.

\section*{Acknowledgements}
We thank David Schiminovich for useful discussions.  EV was supported by the Columbia Prize Postdoctoral Fellowship in the Natural Sciences.  ZH was supported by NASA grant NNX11AE05G.  GLB was supported by National Science Foundation grant 1008134 and NASA grant NNX12AH41G.

\bibliography{IM_paper}

\end{document}